\documentclass[pra,twocolumn,showpacs,superscriptaddress,amsfonts,amsmath]{revtex4}
\usepackage{graphicx}
\def\lsim{\hbox{\lower .8ex\hbox{$\, \buildrel < \over \sim\,$}}}
\def\gsim{\hbox{\lower .8ex\hbox{$\, \buildrel > \over \sim\,$}}}

\begin{document}
%

\newcommand{\sss}{\scriptscriptstyle}

\title{Nonlinear Transport of Bose-Einstein Condensates \\ Through
Waveguides with Disorder}
\author{Tobias Paul}
\affiliation{Institut f{\"u}r Theoretische Physik,
Universit{\"a}t Regensburg, 93040 Regensburg, Germany}
\author{Patricio Leboeuf}
\affiliation{Laboratoire de Physique Th\'eorique et Mod\`eles Statistiques,
Universit\'e Paris Sud, B\^atiment 100, F-91405 Orsay Cedex, France}
\author{Nicolas Pavloff}
\affiliation{Laboratoire de Physique Th\'eorique et Mod\`eles Statistiques,
Universit\'e Paris Sud, B\^atiment 100, F-91405 Orsay Cedex, France}
\author{Klaus Richter}
\affiliation{Institut f{\"u}r Theoretische Physik,
Universit{\"a}t Regensburg, 93040 Regensburg, Germany}
\author{Peter Schlagheck}
\affiliation{Institut f{\"u}r Theoretische Physik,
Universit{\"a}t Regensburg, 93040 Regensburg, Germany}

\begin{abstract}
We study the coherent flow of a guided Bose-Einstein condensate incident over
a disordered region of length $L$. We introduce a model of disordered
potential that originates from magnetic fluctuations inherent to
microfabricated guides. This model allows for analytical and numerical studies
of realistic transport experiments. The repulsive interaction among the
condensate atoms in the beam induces different transport regimes. Below some
critical interaction (or for sufficiently small $L$) a stationary flow is
observed. In this regime, the transmission decreases exponentially with $L$.
For strong interaction (or large $L$), the system displays a transition
towards a time dependent flow with an algebraic decay of the time averaged
transmission.
\end{abstract}


\maketitle

\section{Introduction}\label{sec1}

The extraordinary experimental control achieved over atomic Bose-Einstein
condensates (BEC)  provides new testgrounds for phenomena issued
from many different fields. On the one hand these systems allow to study
extensively nonlinear phenomena such as four-wave mixing \cite{Deng},
propagation of bright \cite{BS} and dark \cite{DS} solitons or the dynamics of
Bloch oscillations in presence of atom-atom interactions \cite{Fallani, Witthaut}.
On the other hand the rapid
progress in this field has lead to a number of fascinating experiments probing
complex condensed matter phenomena, such as the Mott transition in optical
lattices \cite{Greiner}, the creation of vortices \cite{Abo}, the Josephson
effect \cite{Townsend} or the BEC-BCS crossover \cite{Bourdel}. Bose-Einstein
condensates link these two prominent fields of current research in a exciting
and unique way.

Wave mechanical transport in atomic vapors appears as a new direction for
these trans-disciplinary studies that provide deeper insights into transport
phenomena in presence of interaction. Indeed, BEC systems are intrinsically
phase coherent, as are the clean two-dimensional electronic structures studied
in mesoscopic physics at low temperatures. Besides, interaction is much more
simply modeled in BEC systems than the electrostatic electron-electron
potential and its sign (repulsive or attractive) and strength can be tuned
almost at will. The link between matter-wave physics and electronic transport
phenomena became ultimately apparent with the advent of microscopic traps and
waveguides for the atoms, known as {\em atom chips} \cite{FolmanI, Ott,
Haensel}. Related studies include the attempt to generalize Landauer's theory
of conductance to cold atoms \cite{Thywissen}, the {\em atom blockade}
phenomenon in quantum-dot like potentials \cite{Carusotto} as well as
nonlinear resonant transport of Bose Einstein condensates
\cite{PaulSchlagheck}, to mention just a few examples.

A new direction in this context is the transport of Bose-Einstein condensates
through {\em disordered} potentials. A relevant question is to which
extent a Bose-Einstein condensate is subject to Anderson localization
\cite{Anderson58,Fukuyama} in presence of disorder, as well as how this
scenario is affected by the atom-atom interaction. There is a growing interest
in the BEC community for issues related to the behavior of matter waves in
disordered potentials. It started with the observation of ``fragmentation of the
condensate'' over a microchip \cite{corrug}. Nowadays a random potential is
routinely engineered using an optical speckle pattern and
its effects over the expansion of the condensate have
been explored in Refs.~\cite{Lye,Clement}.

In contrast to studies where the condensate is initially at rest, we focus in
the present paper on the effect of disorder on a {\em propagating}
Bose-Einstein condensate. In an adiabatic approximation, the dynamics reduces
to an effective one dimensional (1D) transport problem, this is the so called
``1D mean field regime'' \cite{Menotti}. We furthermore assume that the mean
kinetic energy of the atoms in the condensate is larger than the typical
height of the barriers induced by the disorder potential, i.e., perfect
transmission is expected by classical mechanics. For the sake of concreteness,
we restrict ourselves to one specific type of disorder: the one experienced by
a condensate that is magnetically trapped above a corrugated microchip. To
this end we introduce a model that could be characterized as a ``dirty wire model''
where the current in the microfabricated wire has white noise fluctuations.
This simple model captures most of the caracteristics of the random potentials
observed over corrugated microchips. We point out, however, that our results
are not expected to be sensitive to the particular type of disorder, as long
as the latter is sufficiently smooth and can be characterized by a
well-defined correlation length.

Previous theoretical studies of the effect of disorder on the transmission of
nonlinear waves mainly focussed on attractive interaction and looked for
stationary solutions of the problem (see the review \cite{Gredeskul}). Then,
one has to choose between fixed input and fixed output boundary conditions.
The latter case is less realistic, but simpler to discuss: It leads to
algebraic decay of the transmission \cite{Devillard}. The former case is
complicated by the advent of multistability. However, the results of Knapp {\it
et al.} \cite{Knapp} show that, for short sample size, the mean transmission
is poorly affected by a weak nonlinearity (as compared to the linear case),
whereas for larger samples and stronger nonlinearity, evidences of
delocalization are found.
.

Realistic transport processes of Bose-Einstein condensates are different from the above mentioned studies because they typically involve particles
experiencing {\em repulsive} interactions. We will see below that in this case the assumption of stationarity is not appropriate because
for large  disordered region or strong nonlinearity stationary solutions are
dynamically unstable. In typical experiments the population of a given final state can only be achieved through a time-dependent process (such as the
gradual filling of an initially empty waveguide with matter waves). As a result, if a stationary scattering state is unstable, the transport properties of
the condensate may be unrelated to the transmission coefficient associated with that state, whereas a study of stationary flows might misleadingly give
some weigth to this state (if the transmission is averaged over all possible stationary solutions for instance).

We thus consider a setup that is relevant to experimental realizations and
adapted to this specific transport
scenario explicitely taking into account the possibility of
time-dependent scattering:
a coherent source of atoms emits matter waves that
propagate in the magnetic waveguide and encounter on their path a disorder
region of length $L$. We show that the presence of a repulsive atom-atom
interaction has dramatic effects on the transport properties of the
condensate. As is the case for attractive nonlinearity, Anderson localization
is observed only in the regime of small interaction strengths and sample
lengths. In this regime the transmission deacreases exponentially with $L$
($\propto \exp(-L/L_{loc})$), with a localization length $L_{loc}$ modified by
the interaction. For large sample lengths or strong interaction,
time-dependent scattering processes occur. In contrast to the previous regime,
one observes an Ohmic decrease of the time-averaged transmission ($\propto
L^{-1}$).

The paper is organized as follows. In Sec.~\ref{sec2} we set up the
theoretical framework that is necessary to study transport through
mesoscopic waveguides, introduce an effective one-dimensional
Gross-Pitaevskii equation and present a numerical method that is
particularly suited to study transport processes of Bose-Einstein
condensates in waveguides. In Sec.~\ref{sec3} we introduce a
one-dimensional model for the random magnetic potential along the
center of the waveguide. We will show that a microscopic meandering
of the current in the wire on the atom chip leads to a Lorentzian
correlated random potential. In Sec.~\ref{sec4} we investigate the
regime of weak disorder potentials and give a simple analytic
expression for the condensate wavefunction in the guide. In
Sec.~\ref{sec5} we discuss numerical results for transport through
moderate and strong disorder regions. We consider in particular the scaling
of the transmission with the length of the disorder region. The paper
closes with some concluding remarks. Some technical points are given
in the appendices. In Appendix \ref{sec6} we derive a relation
between the mean transmission and the correlation function of the
disorder potential. In Appendix \ref{sec7} we rederive, using
standard WKB techniques, a result which is obtained heuristically in
the main text.

\section{Transmission through waveguides}\label{sec2}

We consider a coherent beam of Bose-Einstein condensed atoms at zero
temperature, propagating through a cylindrical magnetic waveguide of axis $x$.
The condensate is formed by atoms of mass $m$ which interact {\it via} a
two-body potential characterized by its 3D s-wave scattering length $a_{sc}$.
We consider the case of repulsive effective interaction, i.e., $a_{sc}>0$. The
condensate is confined in the transverse direction by an harmonic potential of
pulsation $\omega_\perp$. This transverse confinement is characterized by the
harmonic oscillator length $a_\perp=(\hbar / m \omega_\perp)^{1/2}$.

In the following we restrict ourself to the ``1D mean field regime''
\cite{Menotti} corresponding to a density range such that
$(a_{sc}/a_\perp)^2\ll n_{1\rm{D}}\, a_{sc} \ll 1 $, where $n_{1\rm{D}}$
denotes a typical order of magnitude of the 1D density $n(x,t)$ of the system.
The first of these inequalities ensures that the system does not get in the
Tonks-Girardeau limit and the second that the transverse wave function is the
ground state of the linear transverse Hamiltonian, see, e.g., the discussion
in Refs.~\cite{Menotti,Petrov}. In this regime the system is described by a 1D
order para\-me\-ter $\psi(x,t)$ (such that $n(x,t)=|\psi(x,t)|^2$) depending
only on the spatial variable $x$ along the guide. $\psi(x,t)$ obeys the 1D
Gross-Pitaevskii equation
\begin{equation} \label{2.3}
 i \hbar \frac{\partial \psi}{\partial t}= \left( -\frac{\hbar^2}{2m}
 \frac{\partial^2 }{\partial x^2} + V(x) +g\,n(x,t) \right) \psi \; ,
\end{equation}
\noindent with $g=2\hbar \omega_{\perp} a_{sc}$
\cite{Olshanii,Jackson,LeboeufPavloff}. $V(x)$ is an effective
one-dimensional potential along the waveguide, to which the condensate is
exposed during the propagation process. We will see in Sec.~\ref{sec3} how it
may originate from irregularities of a wire used for creating the magnetic
confinement.

\begin{figure}[tbp]
\centering
\includegraphics*[width=7.5 cm]{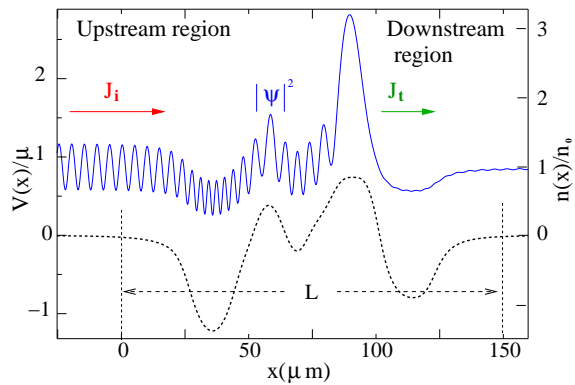}
\caption{\label{fig1} (Color online) A condensed beam with incident current $J_{i}$ can
 populate a stationary scattering state. The solid line shows its longitudinal
 density $n(x)$ (in units of the equilibrium density $n_{0}$). In the
 downstream-region, $\psi$ tends to a plane wave with transmitted current
 $J_{t}$. The dashed line displays the scattering potential $V(x)$ in units of
 the chemical potential $\mu$.}
\end{figure}

In absence of a potential ($V(x)\equiv 0$) the plane wave
\begin{equation} 
\psi(x,t)=\sqrt{n_{0}} \exp(i k x - i \mu t /\hbar)
\end{equation}
is obviously a solution of the Gross-Pitaevskii equation (\ref{2.3}). It
satisfies the dispersion relation
\begin{equation} \label{2.4}
\mu = \frac{m}{2}\frac{J^2}{n_{0}^{2}} + g n_{0} \; ,
\end{equation}
where the particle current is given by $J=n_{0}\hbar k / m$.
Therefore, the chemical potential $\mu$ and the equilibrium constant
density $n_{0}$ of a freely propagating condensate beam are
determined by the current $J$, the wave vector $k$, and the
effective interaction strength $g$. At this point, we mention that
it was demonstrated in \cite{LeboeufPavloff} that Eq.~(\ref{2.4})
exhibits two constant-density solutions: a low-density (supersonic)
one and a high density (subsonic) one, where the transport is
respectively dominated by the kinetic energy or by mutual
interaction of the atoms. 
Both solutions are plane waves of the form $\psi(x)=A e^{ikx}$,
but with different wavevectors $k$ and particle densities $A^2$.
As we are considering rather small condensate densities and large velocities 
in the waveguide, the
supersonic solution will be the relevant one in the context of this
paper.

We now assume the presence of a disorder potential $V(x)$ in the
waveguide which is finite between $x=0$ and $x=L$ and vanishes elsewhere.
In this case, a BEC that is injected into the initially condensate-free
disorder region from the upstream side (i.e. at $x<0$) does in general not freely
propagate to the downstream region (at $x>L$), but undergoes a
scattering process.
In this paper we shall compute transport properties
of a system where a monochromatic beam of condensate with 
well-defined current $J_{i}$ is injected into the disorder region
(see Fig.~\ref{fig1}).
This means, we consider the propagation process in
terms of a so called fixed input problem \cite{Knapp,Gredeskul}. 

Our purpose is now to compute transmission coefficients for the condensate
transport through the disordered region. Furthermore, we shall
investigate to which extent it is possible to populate 
stationary scattering states, i. e. stationary solutions
$\psi(x,t)=\psi(x)\exp(-i\mu t /\hbar)$ 
satisfying the outgoing boundary condition 
$\psi(x)=\sqrt{n_{0}} e^{ikx}$ (with $k>0$) for
$x\rightarrow +\infty$  where $n_0$ is the density associated with the
supersonic solution \cite{bc}.
This question can be addressed by integrating the
time-dependent Gross-Pitaevskii equation (\ref{2.3}) in presence of
a source term that is localized in the upstream region and emits
monochromatic matter waves. Such a source models the coupling of the
waveguide to a reservoir of condensate from which matter waves are
injected into the guide. It has been demonstrated in
\cite{PaulSchlagheck} that this approach is particularly well suited
to compute transmissions for fixed input problems. Additionally, it
allows to determine for a given potential $V(x)$ whether an incident
monochromatic beam populates a stationary scattering state or not.

Hence, we consider now the modified Gross-Pitaevskii equation with
a source which is localized at the position $x_{0}$
in the upstream region,
\begin{eqnarray}\label{2.5}
 i \hbar \frac{\partial\psi(x,t) }{\partial t}&=& \left[ -\frac{\hbar^2}{2m}
 \frac{\partial^2 }{\partial x^2} + V(x) +g\vert\psi(x,t)\vert^2 \right]
\psi(x,t)
 \nonumber\\
 &&+S_{0}\exp(-i \mu t /\hbar)~\delta(x-x_{0}) \; .
\end{eqnarray}
$S_{0}$ is the source amplitude which determines the emitted current. To
understand the functionality of the source term, it is instructive to consider
first solutions of Eq.~(\ref{2.5}) in absence of the potential $V(x)$. In this
case there exists plane wave solutions with constant density $n$. To
demonstrate this, we switch to the Fourier space, where Eq.~(\ref{2.5}) takes
the form (for constant $n$)
\begin{equation} \label{2.6}
\left(\frac{i}{\hbar}\frac{\partial}{\partial t} -\frac{\hbar^2 \, q^2}
{2 m} -g n \right)\tilde\psi(q,t)=S_{0}\,e^{-i q x_0}\,e^{-i\mu t/\hbar}
\; .
\end{equation}
This equation admits a solution of the form
\begin{equation} \label{2.7}
\tilde\psi(q,t)=\frac{S_{0}\,e^{-i q x_0}\,e^{-i\mu t/\hbar}}
{\mu-g n -\hbar^2 q^2/(2 m)}
 \; .
\end{equation}
By transforming back to real space, we find that the source
emits in both directions the monochromatic wave
\begin{equation} \label{2.71}
\psi(x,t)=\frac{S_{0} m}{i k \hbar^2}\;e^{i k \vert x-x_{0}\vert
}\;e^{-i \mu t/\hbar} \; .
\end{equation}
In Eq.~(\ref{2.71}) $k$ is self consistently defined by
$(\hbar k)^2=2 m \left[\mu-g\vert S_{0}\vert^2 m^2/(\hbar^4 k^2)\right]$.
The current emitted by the source can be calculated
by evaluating the quantum mechanical current operator. We find
$J_{i}=\pm\vert S_{0}\vert^2 m/(\hbar^3 k)$; ("$+$" for $x>x_{0}$,
"$-$" for $x<x_{0}$).

We now return to the general case $V(x)\neq 0$. In order to
perform the numerical integration, the wavefunction $\psi(x,t)$ is
expanded on a finite lattice and is propagated in real time domain.
As we are dealing with an open system, artificial backscattering at
the boundaries of the lattice has to be avoided. For that purpose we
impose absorbing boundary conditions that are well suited
for transport problems \cite{Shibata} and can be generalized to
account for weak or moderate nonlinearities \cite{Preparation}. 

As in real experiments we choose as initial condition $\psi(x,t=0) \equiv 0$.
In order to compute the condensate wavefunction we numerically integrate
$\psi(x,t)$ in Eq.~(\ref{2.5}) while adiabatically tuning the source amplitude
$S_{0}$ from $0$ up to a given maximum value that corresponds to a desired
incident current $J_{i}$. This approach simulates a realistic propagation
process, where a coherent Bose-Einstein condensate beam with chemical
potential $\mu$ is injected into the initially empty waveguide from a
reservoir. For comparatively weak nonlinearities a stationary scattering state
of the form $\psi(x,t)=\psi(x)~e^{-i \mu t /\hbar}$, which corresponds to a
supersonic solution in the downstream region, is generally obtained from the
numerical propagation. This stationary wavefunction fulfills the
time-independent Gross-Pitaevskii equation
\begin{equation} \label{2.7a}
\mu~\psi(x)=\left[-\frac{\hbar^2}{2 m}\frac{\partial^2}{\partial
x^2}+V(x)+g\vert \psi(x) \vert^2 \right]\psi(x) \; .
\end{equation}

In contrast to the case of the linear Schr\"odinger equation the
transmission coefficient cannot be computed by simply decomposing
the upstream wavefunction into an incident and reflected part
because the superposition principle is not valid in presence of the
nonlinear term. Such a decomposition is only possible in the limit of
small interaction strengths or small back-reflections \cite{Sinha}.
However, our numerical approach permits nevertheless a straightforward access
to the transmission coefficient also in the nonlinear case.
The latter is evaluated by the ratio of the current $J_{t}$ in presence
of the potential $V(x)$ (i.e., the transmitted current) to the
current $J_{i}$ obtained in absence of $V(x)$ (the incident current
that is emitted by the source).
This approach provides a natural extension of the usual definition
of transmission coefficients in quantum mechanics to the
nonlinear case \cite{PaulSchlagheck}.

In the nonlinear regime, due to dynamical instabilities the
wavefunction $\psi(x,t)$ does not always converge towards a
stationary state but can remain time dependent (cf Sec.~\ref{sec5}).
In that case, the downstream current is no longer constant and
therefore the transmission becomes a function of time. In this case
we simulate the propagation process over a long period 
$\tau$ (ideally $\tau \rightarrow \infty$) and characterize the
transport properties of the guide by means of the time-averaged
transmission
\begin{equation} \label{2.8}
\overline{T} = \lim\limits_{\tau \rightarrow \infty}~\frac{1}{\tau}
\int\limits_{t}^{t+\tau} {T}(t') dt'\; , \qquad (t>0) \; .
\end{equation}
This choice of working with the mean value $\overline{T}$ is
inspired by common experimental setups: the number of condensed
atoms ${\mathcal N}_A$ reaching the downstream region during the time
$\tau$ is ${\mathcal N}_A=\tau ~ \overline{T} ~J_{i}$. This number of
atoms can be determined experimentally, e.g., by use of absorption
spectroscopy.

\section{A simple model of disorder}\label{sec3}
\begin{figure}[tbp] 
\centering
\includegraphics[width=7.5cm,angle=0]{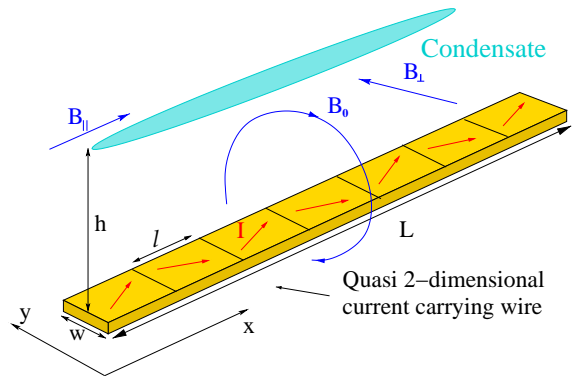}
\caption{\label{fig3} (Color online) Main building block to create a magnetic waveguide on a
 chip: a current flowing in a microfabricated wire and a perpendicular bias
 field form an elongated microtrap. Imperfections in the wire force the
 current to follow a weakly meandering path and generate therefore a magnetic
 disorder potential along the center of the guide.}
\end{figure}

In order to compute transport properties through disordered regions
in magnetic waveguides, it is necessary to introduce an appropriate
model for the static random magnetic potential along the center of
the waveguide. We first briefly recall the basic principle to
generate elongated magnetic waveguides for cold atoms or
condensates. A typical setup that is commonly implemented on atomic
chips is the so called side wire guide \cite{Folman}. 
As sketched in Fig.~\ref{fig3} a circular magnetic field ${\bf B}_{0}$, 
created by an electric current $I$ that flows along a straight
microfabricated quasi 2-dimensional wire and a homogeneous bias field ${\bf B}_{\perp}$
form a minimum of the magnetic field parallel to the wire at
distance $h$. An offset field ${\bf B}_{\vert \vert}$ applied
parallel to the wire reduces losses induced by spin flip processes
near the magnetic field minimum.
For a spatially homogeneous current density in an idealized wire, the magnetic
waveguide is perfectly uniform along its longitudinal axis. 
In reality however, inhomogeneities in the
current density inside the wire have to be taken into account. Such deviations
from an homogeneous current flow can be induced by shape
fluctuations of the wire or impurities inside the metal. These
imperfections cause a magnetic field roughness along the center of
the waveguide that acts as an additional potential and prevents
perfect transmission of condensate beam through the guide
\cite{Aspect}. This additional magnetic field component increases as
the distance to the chip surface diminishes and is expected to
reduce the transmission noticeably.

In the following we consider a steady state current density
${\bf{j}}({\bf r})$ flowing in a thin quasi 2-dimensional metallic
wire. Due to the wire imperfections the current density varies with
the position ${\bf r}$. We decompose ${\bf{j}}({\bf r})$ into a
large constant component ${\bf{j}}_{0}$ flowing parallel to the wire
and a small component ${\bf{\delta j}}({\bf r})$
\begin{equation} \label{3.0}
{\bf j}({\bf r})={j}_{0}~{\bf{e}}_{x}+{\bf{\delta j}}({\bf r}).
\end{equation}
At the center of the waveguide the circular magnetic field ${\bf B}_{0}$
that is generated by ${\bf{j}}_{0}$ cancels with the bias field
${\bf B}_{\perp}$. Hence, the total magnetic field
along the center of the waveguide is given by
\begin{eqnarray} \label{3.0a}
{\bf B}(x,0,h)&=&{B}_{\vert \vert}~{\bf{e}}_{x}+{\bf{\delta  B}}(x,0,h),
\end{eqnarray}
where ${\bf \delta B}=\delta B_{x} {\bf e}_{x} +\delta B_{y} {\bf e}_{y} +
\delta B_{z} {\bf e}_{z}$ is computed from the Biot-Savart law
\begin{equation} \label{3.0b}
{\bf \delta B}= \frac{\mu_{0}}{4 \pi}\int d^3{\bf r}'\;
\frac{ {\bf \delta j}({\bf r}') \times ({\bf r}-{\bf r}')}
{\vert {\bf r}-{\bf r}'\vert^3} \; .
\end{equation}
\begin{figure}[tbp]
\centering
\includegraphics[width=8.5cm,angle=0]{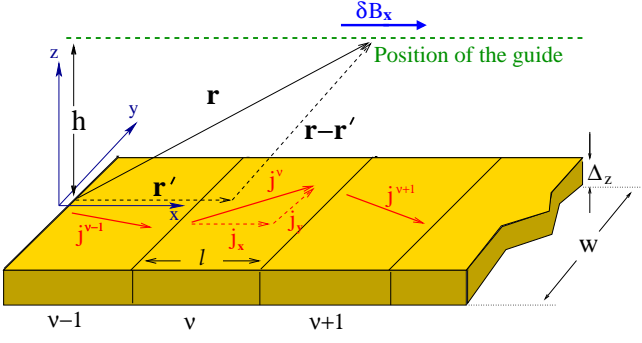}
\caption{\label{fig4} (Color online) Partitioning of the wire in equal blocks of length $l$,
width $w$, and thickness $\Delta_{z}$. For each block we compute an average
current density ${\bf j}^{\nu}$. The current component parallel to the
$y$-direction is at the origin of the magnetic disorder potential along the
center of the waveguide. }
\end{figure}
The effective potential for the atoms is proportional to
the modulus of the magnetic field
\begin{equation} \label{3.0d}
\vert{\bf B}\vert=\sqrt{(B_{\vert\vert}+\delta B_{x})^2+\delta B_{y}^2+
 \delta B_{z}^2}.
\end{equation}
As $\delta {\bf j}$ is supposed to be small, we keep only terms
of first order in $\delta {\bf j}$. This yields the
simple result
\begin{equation} \label{3.0g}
\vert {\bf B} \vert =B_{\vert \vert}+\delta B_{x} \; .
\end{equation}
Hence, within the approximation of small current fluctuations, the
disorder potential along the center of the waveguide
is given by
\begin{equation} \label{3.0h}
V(x)=\mu_{\sss B}~ \delta B_{x}(x,0,h).
\end{equation}

We now consider a quasi 2-dimensional wire of length $L$ in
$x$-direction and width $w$ in $y$-direction. A proper description
of the current density  ${\bf j}({\bf r})$ in the metallic wire
would require an accurate microscopic model for structural
dislocations of the wire as well as its impurities \cite{Aspect, Wang2004}.
In the present work we adopt a more simple and
phenomenological  approach, which is valid if the length scale $l$
on which ${\bf j}({\bf r})$ typically fluctuates is much smaller
than the height $h$ of the waveguide. To this end, we divide the
wire into $\mathcal{N}$ equal blocks of length $l$  width $w$ and
thickness $\Delta_{z}$ (see Fig.~\ref{fig3} and Fig.~\ref{fig4}). For
each block of volume $V$ we compute the average current density
\begin{equation}\label{3.1}
{\bf j}^{\nu}= \frac{1}{V}\int\limits_{-\Delta z/2}^{\Delta z/2}
dz\int\limits_{(\nu-1) l}^{\nu l}dx \int\limits_{-w/2}^{w/2}dy \,
{\bf{j}}(\bf{r}) \; .
\end{equation}
(The index $\nu=1...\mathcal{N}$ labels the blocks and the
corresponding mean current densities ${\bf j}^{\nu}$). The total
electric current along the wire is given by
\begin{equation} \label{3.1a}
I=\int\limits_{-\Delta_{z}/2}^{\Delta_{z}/2}~dz \int\limits_{-w/2}^{w/2}dy~
{\bf j}({\bf r}) \cdot {\bf e}_{x}= w \Delta_{z}~{\bf j}^{\nu} \cdot {\bf e}_{x}.
\end{equation}
Hence, in the usual case of a stationary electric current $I$, the
$x$-component of ${\bf j}^{\nu}$ is given by the constant value $j_0$ of
Eq.~(\ref{3.0}) for all $\nu$, and we have
\begin{equation} \label{3.1b}
{\bf j}^{\nu}=
j_{0}{\bf e}_{x}+\delta j_{y}^{\nu} {\bf e}_{y} + \delta j_{z}^{\nu}.
{\bf e}_{z}
\end{equation}
The thickness $\Delta_{z}$ of the wire is assumed to be much smaller than all
other relevant length scales. We therefore assume $\vert \delta j_{z}^{\nu}\vert
\ll \vert\delta j_{y}^{\nu} \vert$ and neglect the contribution of $\delta
j_{z}^{\nu}$ to the disorder potential in the following. This yields
\begin{equation} \label{3.5}
V(x)= \mu_{\sss B} \sum\limits^{{\mathcal N}}_{\nu=1} \delta {B}^{\nu}_{x}(x,0,h)
\end{equation}
where the magnetic field contribution of the $\nu^{\rm{th}}$ block
at the center of the waveguide is computed from the
Biot-Savart law according to
\begin{eqnarray} \label{3.4}
& & \delta B^{\nu}_{x}=
\frac{\mu_{0}}{4 \pi}\int\limits_{(\nu-1)l}^{\nu l}\!\!\! dx'\!\!
\int\limits_{-\frac{w}{2}}^{\frac{w}{2}}\!\!\! dy'
\frac{\Delta_z\, h\, \delta j^{\nu}_y}
{\left[(x-x')^2+h^2+y'^2\right]^{3/2}}\qquad \;   \\
& & =\frac{\mu_{0}\Delta_z}{2 \pi}\, \delta j^{\nu}_{y}
   \left[ \arctan\left( \frac{w\, u/(2\,h)}{\sqrt{u^2+h^2+w^2/4}}\right)
   \right]^{x-(\nu-1)l}_{x-\nu l}\!\!\!\!\!\!\!\!\!\! .\nonumber
\end{eqnarray}

\begin{figure}[tbp]
\centering
\includegraphics[width=7.5cm,angle=0]{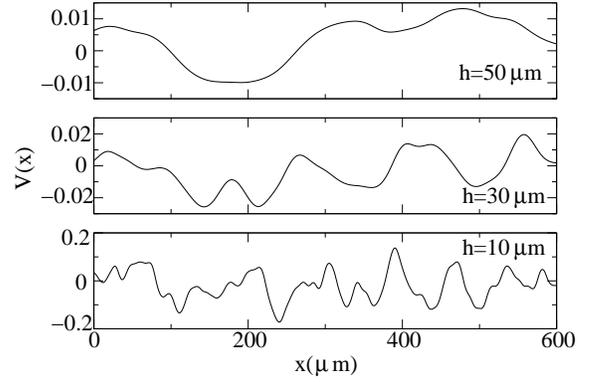}
 \caption{\label{fig5} Numerically computed representative examples for
 disorder realizations at different distances $h$ between the center of the
 guide and the atomic chip surface. The panels show the transition from weak
 to strong disorder, with decreasing distance $h$. The disorder potential is
 given in units of $ \mu_0 \mu_B w \Delta_z \sigma/(2 \pi)$}
\end{figure}
Within the discretization procedure just described, we can introduce
disorder by assuming $\delta j_{y}^{\nu}$ as a random variable,
uniformly distributed in the interval
$[-(3\sigma^2/2l)^{1/2},(3\sigma^2/2l)^{1/2}]$. This assumption corresponds to
a zero average transverse current density 
($\langle \delta j_{y}^{\nu} \rangle=0$) with 
a delta correlation
\begin{equation} \label{3.3a}
\langle \delta j_{y}^{\nu} ~\delta j_{y}^{\nu'} \rangle=
\frac{\sigma^2}{l} \; \delta_{\nu,\nu'} \; ,
\end{equation}
and allows to reach a well defined regime in the limit $l\to 0$. In
that limit $\delta j_{y}^{\nu}$ is replaced by an $x$-dependent
quantity $\delta j_{y}(x)$ verifying \cite{remark}
\begin{equation} \label{3.3b}
\langle \delta j_{y}(x) ~\delta j_{y}(x') \rangle=
\sigma^2 \; \delta(x-x') \; .
\end{equation}
Here, the parameter $\sigma$ fixes a scale for the typical deviation
of the current density from an homogeneous current flow. Since the
fluctuations in $\delta j_y$ are certainly proportional to $j_0=I/(w \Delta_z)$,
we can write $\sigma=j_0 \sqrt{\ell^*}$. Here $\ell^*$ is a characteristic length
depending on the properties of the metallic wire, which can in
principle be found from experimental investigations.

Due to the convolution procedure in Eq.~(\ref{3.4}), the short range
disorder in the electric wire induces a smoothly varying potential
$V(x)$ along the guide. This is clearly visible in Fig.~\ref{fig5}
which shows the disorder potentials that result from three
numerically generated sets of current densities $j_{y}^{\nu}$, at
three different heights $h$ of the waveguide.
The disorder potential is smoother for large distances $h$, and
becomes more rough (and its typical intensity increases) as $h$
diminishes.

\begin{figure}[tbp]
\centering
\includegraphics[width=7.5cm,angle=0]{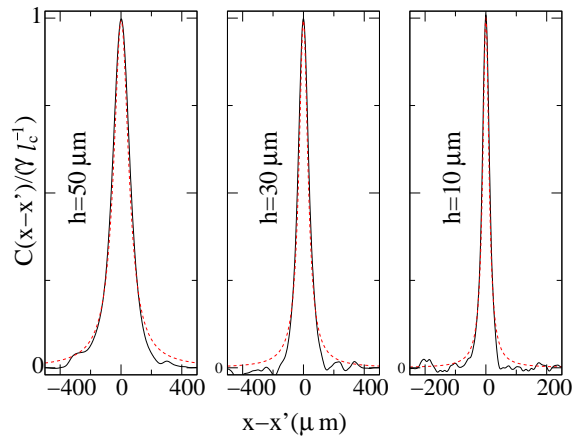}
 \caption{\label{fig6} (Color online) Numerically computed correlation functions for
  different distances $h$ between the center of the guide and the atomic chip
  surface (solid lines). Dashed lines: Fit to a Lorentzian curve. }
\end{figure}

One has $\langle V(x)\rangle =0$ and it is
appropriate to characterize the random potential by studying
 the correlation function 
\begin{equation} 
C(x-x')=\langle V(x) ~V(x') \rangle .
\end{equation}
In Fig.~\ref{fig6} we show results for $C(x-x')$ at different
heights $h$. The correlation function is computed numerically by
averaging over a large number of different disorder realizations. We
find that it can be fitted with good accuracy by a Lorentzian
curve
\begin{equation} \label{3.7}
C(x-x')\simeq \frac{\gamma \, l_{c}}{l_{c}^2+(x-x')^2}\; .
\end{equation}
This allows to extract the correlation length $l_{c}$ and to establish an
empirical relation between the height $h$ and $l_{c}$. In the regime where the
width $w$ of the wire and the discretization length $l$ are of the same order,
we find that the correlation length depends linearly on the distance between
wire and waveguide, $l_c \simeq s\, h$, with a proportionality constant $s$
that varies between 1 and 2. For the experimentally relevant case of $w =
4\, \mu$m (a wire of this size has been realized by the T\"ubingen group, see
Ref. \cite{Zimmermann2}) we find $s=1.2$.
Theoretically, this result may be understood as follows. In the
continuous limit $l\rightarrow 0$
(and in the idealized case of an infinitely long wire),
Eq.~(\ref{3.4}) takes, in the regime $w \gg h$, the particularly simple form 
\begin{equation} \label{3.8}
V(x)=\frac{\mu_{\sss B}\mu_{0}}{2 \pi}
\int\limits_{-\infty}^{+\infty}dx'
\frac{ h\, \Delta_z\, \delta j_{y}(x')}{(x-x')^2+h^2} \; . 
\end{equation}
In this case the disorder potential is exactly Lorentz-correlated with
$l_c=2\,h$ and $\gamma=\pi^{-1}(\mu_0\mu_{\sss B}\Delta_z\sigma/2)^2$. In the
opposite regime $w\ll h$, the correlation function $C(x-x')$ cannot be
computed analytically, but its Fourier transform,
$C_q=\int_{-\infty}^{+\infty}\exp(-i\,q\,x) C(x)dx$, can be calculated: One
obtains
\begin{equation}\label{eq2}
C_q=\left(\frac{\mu_0\,\mu_{\sss B}\,w\,\Delta_z\,\sigma}{2\,\pi}\right)^2\,
\Big[q\, K_1(qh)\Big]^2 \; ,
\end{equation}
\noindent where $K_1$ is the modified Bessel function of the first
kind \cite{Abramowitz}. $C_q$ as given in Eq.~(\ref{eq2}) is not very
different from the Fourier transform of a Lorentzian (a decreasing
exponential), and this is the reason why $C(x-x')$ can be fitted
reasonably well by a Lorentzian also in the regime $h\gg w$. To find
a sensible Lorentzian fit, one can for instance try to reproduce
$C(0)$ and $C''(0)$ obtained from Eq.~(\ref{eq2}) with the parameters
$\gamma$ and $l_c$ of Eq. (\ref{3.7}): $C(0)=\gamma/l_c$ and
$C''(0)=-2\gamma/l_c^3$. This leads to
\begin{equation}\label{eq3}
\frac{l_c}{h}=\left[
\frac{2\,\int\limits_{0}^{+\infty}t^2 K_1^2(t)\,dt}
     {\int\limits_{0}^{+\infty}t^4 K_1^2(t)\,dt}
\right]^{1/2} \simeq 1.46 \; .
\end{equation}
Thus, again in this limit, we find that $l_c$ is of the type
$l_c\simeq s\,h$. The important outcome of this discussion is that,
for the continuous model ($l\to 0$) -- in both limits $w\ll h$ and
$w\gg h$ -- and also in the numerical realizations of the disorder
with a finite grid $l$, we obtain a random potential which is
Lorentz-like correlated, with a correlation length $l_c$ that is
proportional to the height $h$ of the trap above the chip, and with a
proportionality constant of the order $1-2$. This is confirmed
experimentally by the detailed studies presented in Ref.
\cite{Aspect}.

The model introduced in this section, where the disorder potential originates
from white-noise correlated fluctuating currents (see
Eqs.~(\ref{3.3a},\ref{3.3b})) corresponds physically to a ``dirty wire
model'', in the sense that the very erratic random current density
(\ref{3.3b}) can be considered as originating from the presence of impurities
in the wire. We note that the white-noise current correlations lead to a
disorder potential whose typical amplitude (for the continuous limit $l\to 0$,
when $h\gg w$) varies as $\langle V^2(x) \rangle^{1/2}\propto I\,h^{-3/2}$,
different from the experimental finding $I\, h^{-2.2}$ of Kraft {\it et al.}
\cite{Kraft}. In contrast, the model of ``clean wire with corrugated
boundaries'' introduced in \cite{Wang2004} and developed in \cite{Aspect}
yields {\it in the case of a white-noise correlated boundary roughness} to a
dependence of the form $I\, h^{-5/2}$, in closer agreement with the
experimental findings of the T\"ubingen group published in Ref. \cite{Kraft}.
Note however, that the experimental results of the Orsay group \cite{Aspect}
point to a boundary roughness which is not white-noise correlated, and a
typical disorder potential which decreases less rapidly than $h^{5/2}$, as
found in the present study. Also the correlation function issued from the
dirty wire model is in better agreement with the experimental one determined
in Ref. \cite{Aspect}, which differs from the one resulting from a wire with a
white noise disordered boundary (that has a correlation function verifying
$C_{q=0}=0$ \cite{Wang2004}). It thus appears that the simple dirty wire model
introduced in the present section allows to construct a disordered potential
$V(x)$ that captures most of the characteristics of the micro fabricated
magnetic guides.

\section{Weak disorder}\label{sec4}

In this section we investigate the regime of weak disorder potentials and
derive simple relations between the condensate density and the disorder
potential $V(x)$. 
{\em Weak disorder} means in this context that the propagation of the condensate
is only marginally affected by the scattering region.
This implies that the kinetic energy 
per particle must be much larger than the typical intensity of the
disorder potential (which can be estimated for instance by the standard
deviation $\langle V^2(x) \rangle^{1/2}$).
We shall argue below that a secondary criterion is necessary to
characterize this regime, namely that the length of the disordered region 
is small compared to the characteristic length scale $L_d$ (to be defined below)
typical for the decrease of the transmission.

First, we rewrite the Gross-Pitaevskii equation~(\ref{2.3}) in the well
known form of hydrodynamic equations
\begin{equation} \label{4.1a}
\frac{\partial }{\partial t}n=-\frac{\partial}{\partial x}(n u) \; 
\end{equation}
and
\begin{equation}\label{4.1b}
m\frac{\partial u }{\partial t}=\frac{\partial}{\partial x}  \left[
\frac{\hbar^2}{2m {n}^{1/2}}
\frac{\partial^{2}{n}^{1/2}}{\partial x^{2}} - \frac{m u^2}{2}
-V(x)-gn  \right]\, ,
\end{equation}
where ${u}$ is the condensate velocity.
In the case of a stationary state we have $\partial_{t} n=0$
and $\partial_{t} u=0$ from which it follows that the current
$J = n u$ is constant.
Integration of Eq.~(\ref{4.1b}) then yields
\begin{equation} \label{4.2}
\mu = V(x)+gn+\frac{J^2}{2 m n^2}-
\frac{\hbar^2}{2m {n}^{1/2}}
\frac{\partial^{2}{n}^{1/2}}{\partial x^{2}}.
\end{equation}
This is the time-independent Gross-Pitaevskii equation for
a current carrying scattering state.
In the downstream region an outgoing plane wave $\psi(x)=\sqrt{n_{0}} e^{ikx}$
is expected. The equilibrium density $n_{0}$ coincides with the
supersonic solution of the dispersion relation~(\ref{2.4}).
Defining $\rho(x)\equiv n(x)/n_{0}$ and
$v(x)\equiv 2 m V(x) / (\hbar^2 k^2)$ one may rewrite Eq.~(\ref{4.2}) in a
dimensionless form:
\begin{equation} \label{4.2a}
-\frac{1}{\rho^{1/2}}\frac{\partial^2 \rho^{1/2}}{\partial x^2}
+\frac{\rho-1}{\xi^2}
+k^2\left[\frac{1}{\rho^2}-1+v(x)\right]=0 \; .
\end{equation}
In this expression we made use of the dispersion relation Eq.~(\ref{2.4}) and
expressed $J=n_{0} \hbar k / m$ in terms of the downstream density $n_{0}$ and
of the outgoing wave vector $k$. The quantity
$\hbar^2 k^2/(2m)=\mu - g n_{0}$ is the kinetic energy of the
outgoing plane wave with equilibrium density $n_{0}$.
$\xi=\hbar/\sqrt{2 m n_{0} g}$ is the condensate healing length.

In order to find perturbative solutions of Eq.~(\ref{4.2a}) for
$v(x)\ll 1$, we insert the ansatz $\rho(x)=1+\delta \rho(x)$ into
Eq.~(\ref{4.2a}) and keep only terms that are linear in $\delta
\rho(x)$:
\begin{equation} \label{4.2b}
\frac{\partial^2}{\partial x^2}\delta\rho(x)+4 \kappa^2
\delta\rho(x) = 2 k^2~v(x) \; ,
\end{equation}
\noindent where
\begin{equation} \label{4.2z}
\kappa = k \sqrt{1-\frac{1}{2 \xi^2 k^2}} \; .
\end{equation}
The solution of Eq.~(\ref{4.2b}) in presence of the downstream
boundary conditions $\delta \rho(L)=0$, $\delta \rho'(L)=0$
(flat downstream density) is \cite{harmosc}
\begin{eqnarray} \label{4.2c}
\delta \rho(x)&=&\frac{k^2}{\kappa}\int\limits_{x}^{L}
\sin\Big[2\kappa(x'-x)\Big] \, v(x')\,d x'\; , \nonumber\\
\delta \rho'(x)&=&-2 k^2\int\limits_{x}^{L}
\cos \Big[2\kappa(x'-x)\Big]\, v(x')\,d x' \; .
\end{eqnarray}
This implies that the density profile in the upstream
region ($x<0$) deduced from the linearized Eq.~(\ref{4.2b})
is of the form $n(x)=n_{0}\left[1+\delta \rho(x) \right]$ with
\begin{equation} \label{4.2d}
\delta \rho(x)=
\delta \overline{\rho}\,  \cos(2 \kappa x + \theta) \; .
\end{equation}
The amplitude $\delta \overline{\rho}$ and the phase factor $\theta$ in
Eq.~(\ref{4.2d}) are determined by the disorder potential $V(x)$ via Eq.~(\ref{4.2c}).
The modified wave number $\kappa$ fixes the period of the density
oscillations.

As we are obviously in the regime of small back-reflections
we adopt the method of Ref.~\cite{Sinha} to determine the
transmission coefficient in an approximative way.
To this end we make the ansatz
$n(x)=|\psi_{\mbox{\tiny{inc}}}(x)+\psi_{\mbox{\tiny{ref}}}(x)|^2$
with
\begin{eqnarray}\label{up3}
\psi_{\mbox{\tiny{inc}}}(x) & = & a \,\exp\{i \kappa x\}\; ,\nonumber \\
\psi_{\mbox{\tiny{ref}}}(x) & = & b \,\exp\{i (\kappa x+\theta) \} \; .
\end{eqnarray}
Comparing the corresponding density profile with (\ref{4.2d}), one
obtains the following expressions for the amplitudes $a$ and $b$
\cite{achtung}:
\begin{eqnarray}\label{up4}
\frac{a^2}{n_0} & = & 1-\frac{1}{4}\delta\overline{\rho}^{\,2} +
{\cal O}(\delta\overline{\rho}^{\,4})  \; , \nonumber \\
\frac{b^2}{n_0} & = &
\frac{1}{4}\delta\overline{\rho}^{\, 2}+
{\cal O}(\delta\overline{\rho}^{\, 4}) \; .
\end{eqnarray}
It was pointed out in Ref.~\cite{Sinha}, and numerically confirmed for single
and double barrier potentials \cite{Preparation}, that
$\psi_{\mbox{\tiny{ref}}}$ can be approximately identified with the reflected
component of the condensate in the case of almost perfect transmission. This
corresponds to a reflexion coefficient $R=b^2/a^2=\frac{1}{4} \,
\delta\overline{\rho}^{\, 2}+{\cal O} (\delta\overline\rho^{\,4})$ and to a
transmission coefficient which can be expressed (using Eq.~(\ref{4.2d})) as
\begin{equation}\label{4.2h}
T=1-\frac{1}{4}\delta \overline{\rho}^{\,2} =
1-\frac{1}{4} \left\{
\left[\delta \rho(0)\right]^2+
\frac{1}{4~\kappa^2} \left[\delta \rho'(0)\right]^2
\right\} \; .
\end{equation}
In this final expression $\delta \rho(0)$ and $\delta \rho'(0)$ are related to
the disordered potential by means of Eq.~(\ref{4.2c}).
Therefore, determining the transmission $T$ for a given potential $V(x)$
amounts to compute the integrals  Eq.~(\ref{4.2c}).

As shown in Appendix~\ref{sec6} the above procedure allows to
 determine the disorder average $\langle T \rangle$ from the knowledge of the
 correlation function of the disorder potential. For the relevant case of a
 Lorentzian correlation (of the form (\ref{3.7})) we obtain
\begin{equation} \label{4.2i}
\langle T \rangle=1-\frac{L}{L_{d}} \; ,
\end{equation}
where
\begin{equation} \label{4.2i1}
L_{d}=\frac{\hbar^4 \kappa^2}{\pi \gamma  m^2}e^{2 \kappa l_{c}}
\end{equation}
is the characteristic length scale for the decay of the
transmission. We recall here that the above analysis is only valid
in the regime $\delta\overline\rho \ll 1$, i.e., the linear decrease
of $\langle T \rangle$ in Eq.~(\ref{4.2i}) is only valid for $L\ll
L_{d}$. Thus, we have to refine our definition of weak disorder: not
only the intensity of the potential should be small, but also the
length of the disordered region should not exceed the value $L_{d}$.

As we see from expression (\ref{4.2i1}), the effect of the atom-atom
interaction is entirely contained within the modified wave number $\kappa$
(Eq.~(\ref{4.2z})) which describes the period of the upstream density
oscillations. For repulsive atom-atom interactions, we have $\kappa < k$,
which implies that the mean transmission is reduced compared to the
noninteracting case. This behavior is indeed well confirmed by numerical
computations based on the approach presented in Sec.~\ref{sec5}. This
interaction-induced decrease of the transmission was already observed in
Ref.~\cite{Sinha} and interpreted as a lack of kinetic energy compared to the
interaction-free case.

In the limit of very small correlation lengths, i.e. $\kappa l_c\ll 1$, the
disorder potential can be approximated by a white noise potential with
correlation function $C(x-x')\simeq \gamma \pi\delta(x-x')$. Considering the
noninteracting case ($\kappa = k$) we recover in this regime the well-known
expression $L_d=L_{loc} \equiv (\hbar^4 k^2)/(\pi m^2 \gamma)$ for the
localization length of $\delta$-correlated disorder potentials (see, e.g.,
Ref. \cite{Pastur}).

The opposite limit $\kappa l_{c} \gg 1 $ can be considered as the
{\em semiclassical} regime, where the de Broglie wavelength $\lambda
\equiv 2 \pi/k$ of the condensate is much smaller than the
correlation length $l_{c}$ of the disorder potential. In this
regime, the length scale $L_d$ is dominated by the exponential
prefactor $\exp(2 \kappa l_{c})$, and the deviations from perfect
transmission $\langle T\rangle \equiv 1$ vanish exponentially fast
with increasing ratio $\kappa l_{c}$. The semiclassical condition
$\kappa l_{c} \gg 1 $ furthermore allows to derive a simple
analytical expression for the density $n(x)$ throughout the
scattering region. We start from the zeroth-order solution
$n(x)\equiv n_{0}$ valid for $V\equiv 0$. Then, for given $\mu$ and
$J$ the density $n_{0}$ can be obtained by iteratively solving the
selfconsistent equation (strictly equivalent to (\ref{2.4}))
\begin{equation} \label{4.3}
n_0=\sqrt{\frac{m}{2}} \, J \, \left[ \mu- g n_0
\right]^{-\frac{1}{2}} \; ,
\end{equation}
\noindent starting, e.g., with $n_0=J \sqrt{m /(2 \mu)}$. This
procedure guarantees convergence toward the supersonic solution
of Eq.~(\ref{2.4}).

The natural generalization of Eq.~(\ref{4.3}) to the case of a small
but non-vanishing potential $V(x)$ is obtained by using Eq.~(\ref{4.2}) instead
of Eq.~(\ref{2.4}). This yields
\begin{equation} \label{4.4}
n=n_{0} \left[-\frac{V(x)}{\mu-g n_{0}} +\frac{\mu-g n+
\frac{\hbar^2}{2 m \sqrt{n}}\frac{\partial^2}{\partial x^2} \sqrt{n}}
{\mu-g n_{0}}\right]^{-\frac{1}{2}} \; ,
\end{equation}
where the current $J$ was substituted by means of the dispersion
relation~(\ref{2.4}). We shall now find approximate solutions of this
selfconsistent equation in the case of weak disorder, i.e. $\vert v(x) \vert
\ll 1$, where the typical value of $V$ is much smaller
than $\mu- g n_0$ which is the kinetic energy per particle.
We emphasize that this does not imply that the nonlinear term $g
n_{0}$ should be small. 

The zeroth order solution of Eq.~(\ref{4.4}) is
simply the constant equilibrium density $n_{0}$. Resubstituting this constant
solution into the recursive equations yields the first-order solution for the
condensate density
\begin{equation} \label{4.5}
 n^{(1)}(x)=\frac{n_{0}}{\sqrt{1-v(x)}} \; .
\end{equation}
Corrections to this first-order expression particularly arise from
the quantum pressure term
$\frac{\hbar^2}{2~m~\sqrt{n}}\frac{\partial^2}{\partial x^2}
\sqrt{n}$. It can be shown, however, that the latter is suppressed
by a factor $\sim 1/(k l_{c})^2$ as compared to the kinetic energy
$\hbar^2 k^2/(2 m)$ when $n^{(1)}(x)$ is resubstituted in
Eq.~(\ref{4.4}). In the semiclassical regime $k l_{c} \gg 1$, the
quantum pressure term becomes negligible, and the expression
(\ref{4.5}) represents a very good approximation to the actual
density of the condensate in the scattering region. We
show in Appendix~\ref{sec7} that the result~(\ref{4.5})
can be derived in a way that is directly analogous
to the semiclassical WKB approach. 

\begin{figure}[tbp]
\centering
\includegraphics[width=7.5cm,angle=0]{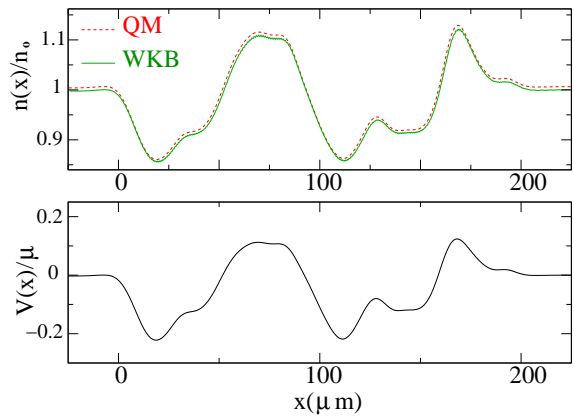}
 \caption{\label{fig7} (Color online) The upper panel displays a comparison of the first
order solution (WKB, Eq. (\ref{4.5})) with a numerically computed solution
(QM) of the Gross-Pitaevskii equation for a weak disorder potential $V(x)$
(shown in the lower panel). The correlation length is $l_{c}=30\,\mu$m, the
wavelength is $\lambda=3\,\mu$m. The ratio between interaction and kinetic
energy in the incident beam is $E_{int}/E_{kin}=1/10$.}
\end{figure}

This result is illustrated in Fig.~\ref{fig7} whose lower panel
shows a random potential generated with the method presented in the
previous section. In the upper panel we compare the result of
Eq.~(\ref{4.5}) with an exact, i.e. numerically computed solution of the
Gross-Pitaevskii equation. Excellent agreement between the first
order solution and the exact solution is found. We note here that it
is quite natural to find that the density profile mirrors the
potential because we are dealing with current carrying states:
the relation (\ref{4.2}) (in absence of quantum correction)
$\mu= m u^2 / 2 +V(x) + g n$ predicts that the condensate velocity
becomes minimal close to the maxima of the disorder potential. It
follows then from the continuity equation $J= n u = \text{const}$
that the density $n$ assumes its maxima when the velocity becomes
minimal.

It is instructive to realize that classically forbidden
back-reflections can be taken into account by inserting the ansatz
$n(x)=n^{(1)}(x)+\delta n(x)$ into Eq.~(\ref{4.2}) and linearizing
the resulting equation for small $\delta n(x) / n_{0}$. To the
lowest non vanishing order in $v$, we again obtain the
result~(\ref{4.2i}) for the mean transmission.

Finally we consider experimental realizations of waveguides on atom chips.
Typical distances $h$ between the chip surface and the guide are in the range
$20$ - $100 \, \mu$m. Typical disorder correlation lengths are of the same
order as $h$. In recent transport experiments \cite{Zimmermann1} the velocity
of propagating $^{87}$Rb condensates is of the order of a few millimeters per
second, resulting to a mean wavelength of a few micrometers. This corresponds
to the regime $\kappa\,l_c\gg 1$ with -- from (\ref{4.2i1}) -- a very large value
of $L_d$. Hence the regime of weak disorder is presently the most relevant
one: the kinetic energy is much larger than the typical intensity of the
disordered potential and $L_d$ is large compared to the typical length of the 
disordered region:
one thus expects almost perfect transmission.

\section{Moderate and strong disorder}\label{sec5}

In Sec.~\ref{sec4} we focused on weak disorder potentials, in the limit of
small reflection. The analysis was done in the regime $\mu\gg
\langle V^2(x)\rangle^{1/2}$ and $L\ll
L_d$. In the present section we still partially fulfill the first of these
inequalities, but drop the second one. We will see that the
behavior of the system is quite different, ranging from a regime of
localization (in the limit of weak interaction), to a time dependent behavior
for larger interaction, with a power law decay of the time-averaged transmission.

First, we shall discuss some elementary differences between the scattering
problem in linear quantum mechanics and the nonlinear Gross-Pitaevskii
equation. In linear quantum mechanics $(g=0)$ one finds for any scattering
potential a unique stationary scattering state that is dynamically stable, and
the associated transmission coefficient $T$ relates the constant incident
current $J_{i}$ one-to-one with the transmitted current $J_{t}$. For the
nonlinear Gross-Pitaevskii equation the transmission $T$ depends on the
density of the propagated condensate and thereby on the current. Additionally,
the phenomenon of multistability may arise.
This means that for a given incident current $J_{i}$ two or more
scattering states with different transmissions can coexist. 

In principle all stationary scattering states that are associated with a given
incident current $J_{i}$ can be found by integrating the time-independent
Gross-Pitaevskii equation~(\ref{2.7a}) from the downstream to the upstream
region. A systematic variation of the downstream current $J_{t}$ allows to
select the desired states. This procedure, however, does not reveal any
information about their dynamical stability properties, which are crucial for
answering the question whether an incident condensate beam populates a
stationary scattering state or not. For instance, in the case of coherent
condensate transport through a double barrier potential, three possible
scattering states are expected close to the resonances, but only one of them
is dynamically stable \cite{PaulSchlagheck}. Here the advantage of integrating
the time-dependent Gross-Pitaevskii equation becomes apparent:  If this
integration converges to a stationary scattering state we know automatically
that this state is dynamically stable (otherwise small numerical deviations
would exponentially increase with propagation time).

We consider an ensemble of $N$ disorder realizations with randomly varying
sample lengths $L$ that are uniformly distributed between $0$ and a maximal
sample length. For each realization (labeled with index $\alpha$) we
numerically compute the time evolution of the wavefunction and extract either
the time-independent transmission $T_{\alpha}$ (if $\psi(x,t)$ converges to a
stationary state) or the time averaged transmission $\overline T _{\alpha}$
(if $\psi(x,t)$ remains time-dependent). For the sake of definiteness and due
to its experimental relevance we consider the propagation of condensed
$^{87}$Rb atoms (whose scattering length is $a_{sc}=5.77$ nm). Our numerical
computations were performed for a guide with radial trapping frequency
$\omega_{\perp}=2 \pi\times 100\,$ s$^{-1}$ (oscillator length $a_{\perp}=1\,
\mu$m). The disorder is generated as in the previous section. The regime of
strong disorder is reached by choosing a rather short distance $h=5\, \mu $m
between the center of the guide and the chip surface, which corresponds to a
correlation length $l_{c}= 6\, \mu $m. In order to avoid excitations of the
condensate
into higher transversal modes we adjust the standard deviation of
the potential (which is a measure of the mean potential height) to
$\langle V^2(x) \rangle^{1/2}\simeq 0.12 \, \hbar \omega_{\perp}$.
In {\em all} the following numerical calculations we consider an incident monochromatic
beam with current $J_{i}=10^{3}$ atoms per second
and wavelength $\lambda=10\, \mu$m.
Then the chemical potential is $\mu = 0.25 \, \hbar \omega_{\perp}$
(in the linear case the chemical potential takes the slightly different
value $\mu = 0.23 \, \hbar \omega_{\perp}$).

\begin{figure}[tbp]
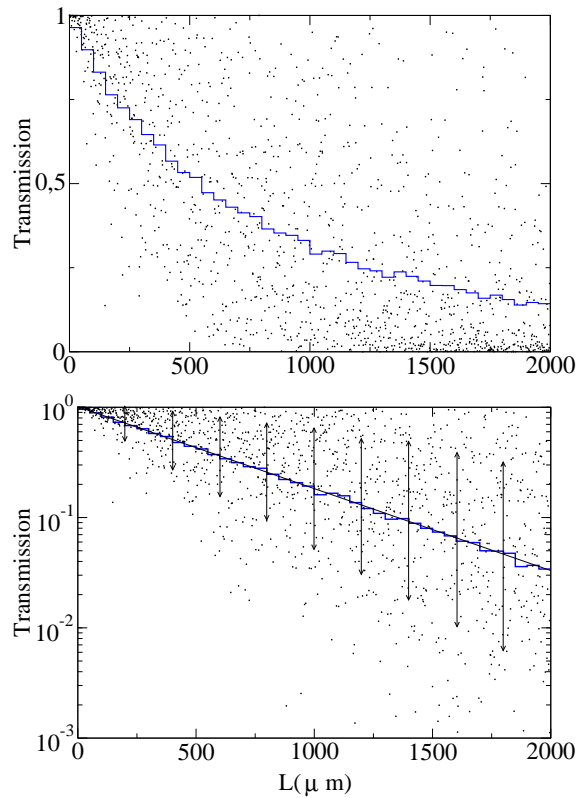

\centering
\includegraphics[width=7.5cm,angle=0]{pics/lin_arit.eps}
\\[2mm]
\includegraphics[width=7.5cm,angle=0]{pics/Lin_geo.eps}
    \caption{\label{fig8} (Color online) Transmission through a disordered sample as a
    function of sample length $L$ for the non-interacting case. Each point
    corresponds to a different realization of the disordered
    potential. Upper panel: arithmetically averaged transmission (blue
    staircase function). Lower panel: The geometric averaged
    transmission (blue staircase function) decreases exponentially with $L$,
    as revealed by the fit with $L_{loc}=586 \,\mu$m (straight black line).
    The arrows mark the logarithmic standard deviation.}
  \end{figure}

It is instructive to focus first on the linear case ($g=0$) that has already
been extensively investigated in the context of localization
theory \cite{Pastur,Gredeskul}. In the localized regime the
transmission decays exponentially with the system length $L$, i.e.,
$\langle T \rangle=\exp(-L/L_{loc})$ where $L_{loc}$ is the so
called localization length. The points in the upper panel of
Fig.~\ref{fig8} mark for each disorder realization the associated
transmission $T_{\alpha}(L)$ as a function of the sample length $L$.
To extract from these data a characteristic scaling law for the
$L$-dependence of the transmission we divide $L$ into equal
intervals of length $\Delta L \ll L$. We then compute the mean
transmission at sample length $L$ by summing up all the values
$T_{\alpha}$ corresponding to a sample length lying in the interval
of width $\Delta L$ centered at $L$:
\begin{equation} \label{5.1}
\langle T \rangle_{a}(L)=
\frac{1}{N_{L}} \sum\limits_{\alpha} T_{\alpha}(L'), \quad
L-\frac{\Delta L}{2}<L'<L+\frac{\Delta L}{2}
\end{equation}
$N_{L}$ is the number of samples in the interval under consideration. 

The step function in Fig.~\ref{fig8} shows the decrease of $\langle T
\rangle_{a} $ for $30000$ disorder realizations and $\Delta L = 50$ (for the
sake of clarity we show only $2000$ points in the plot). In the context of
localization theory it is convenient to investigate scaling laws by means of
the geometrically averaged transmission
\begin{equation} \label{5.2}
\langle T \rangle_{g} = e^{ \langle \ln(T) \rangle }, \quad
\langle \ln(T) \rangle= \frac{1}{N_{L}} \sum\limits_{\alpha} \ln(T_{\alpha}(L')).
\end{equation}
because, contrary to $\langle T \rangle_{a}$, the average $\langle \ln(T)
\rangle$ is a self-averaging quantity of the system \cite{Pastur, Thouless}.
The lower panel of Fig.~\ref{fig8} shows $\langle T \rangle_{g}$ which follows
clearly an exponential law. This is a clear evidence for the appearance of
localization. We can extract the localization length that is here
$L_{loc}=586\, \mu$m. We note the wide spread of the data points around their
average. This spread is quantified by the logarithmic standard deviation
\begin{equation} \label{5.3}
{\Delta \ln(T)}= \left[\sqrt{\frac{1}{N_{L}}
\sum\limits_{\alpha}
\Big[ \ln(T_{\alpha})- \langle \ln(T) \rangle \Big]^2}\; \right] \; ,
\end{equation}
which is shown as arrows in the lower panel of in Fig.~\ref{fig8}. We find an
almost linear increase of ${\Delta \ln(T)}$ with the sample length.

\begin{figure}[tbp]
\centering
\includegraphics[width=7.5cm,angle=0]{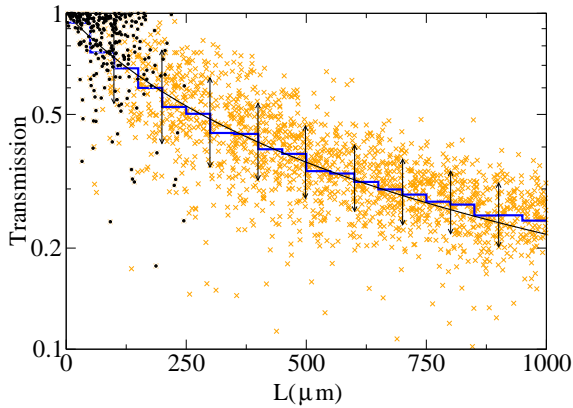}
    \caption{\label{fig9} (Color online) Transmissions through an ensemble of disorder
    realizations for a moderate nonlinearity $E_{int}/E_{kin}=1/10$;  
    The characteristics of the incident beam are given in the main text.
    A transition
    from a time-independent to a time-dependent regime is observed at critical length
    $L*\simeq 125\,\mu$ m. The black dots represent the transmissions in the
    time-independent regime and the (orange online) crosses are the time averaged 
    transmissions in the time dependent regime. The staircase function is the 
    geometrically averaged transmission. It is well approximated    
    by the algebraic scaling law $L_{0}/(L+L_{0})$ (smooth
    solid line) with $L_{0}=287\,\mu$m.}
  \end{figure}

Is the conventional localization scenario, with the characteristic exponential
decrease of the transmission \cite{Anderson58, Fukuyama}, still valid in the
case of interacting atoms? To address this question we now calculate the
transport in presence of a moderate nonlinearity where the ratio of
interaction and kinetic energy in the incident beam is $E_{int}/E_{kin}\simeq
1/10$. Contrarily to the linear case, time-dependent behavior becomes now a
dominant feature as shown in Fig.~\ref{fig9}. We find that dynamical stable
scattering states (black dots in Fig.~\ref{fig9}) are only populated for
sample lengths that are smaller than a critical length $L^*$ which is here of
order of 125 $\mu$m. For samples with
length $L \gsim L^*$ we find a crossover region where time-dependent behavior
sets in and convergency to a stationary state is only achieved for a certain
fraction of disorder samples. $\psi(x,t)$ remains time-dependent (orange crosses)
for all samples when we reach the regime where $L$ is notedly larger
than $L^*$. In the time-dependent case the data points display the time
averaged transmissions $\overline T _{\alpha}$ (\ref{2.8}).

In order to extract a scaling law from our data,
we compute the ensemble-averaged transmission (in the
time-dependent cases $T_{\alpha}$ in Eqs.~(\ref{5.1}-\ref{5.3}) is
replaced by $\overline T _{\alpha}$). We find that the geometric
averaged transmission $\langle T \rangle_{g}$ (step function in
Fig.~\ref{fig9}) decreases inversely with the sample length $L$ and
is well approximated by
the algebraic function (smooth line in Fig.~\ref{fig9})
\begin{equation} \label{5.4}
\langle T \rangle_{g} = \frac{L_{0}}{L+L_{0}}
\end{equation}
with the decay length $L_{0}$. Such a
scaling law is characteristic for transport in systems with loss of phase
coherence between the single scattering events. Indeed, if one
considers a series of successive scatterers and calculates the transmission by
neglecting all interference effects one derives exactly the scaling law of
Eq.~(\ref{5.4}) \cite{Datta,Berry-Klein}. Such an ohmic behaviour is
observed for electron transport through mesoscopic metal structures in the limit
of small dephasing lengths \cite{Datta,Ferry}.

Another striking feature is the distribution of the data points in
Fig.~\ref{fig9}. Contrarily to the linear case, this distribution is now
clearly restricted to the neighborhood of the average transmission and the
standard deviation ${\Delta T}(L)$ decreases for long sample lengths $L$.
Hence, in the regime of large lengths one expects to find the ${\overline T}
_{\alpha}$'s in a narrow interval centered around the
averaged transmission. Loosely speaking, ${\overline T} _{\alpha}$ becomes
more or less sample independent. For the sake of completeness we mention that
ideally the time averages ${\overline T} _{\alpha}$ should be computed for an
infinitely long period. Of course this cannot be done numerically, but we
verified that the averaged transmission and the standard deviation do not
change if we increase in (\ref{2.8}) the averaging time window from $\tau$ to
$2 \tau$ and $3 \tau$.

\begin{figure}[tbp]
\centering
\includegraphics[width=8.99cm,angle=0]{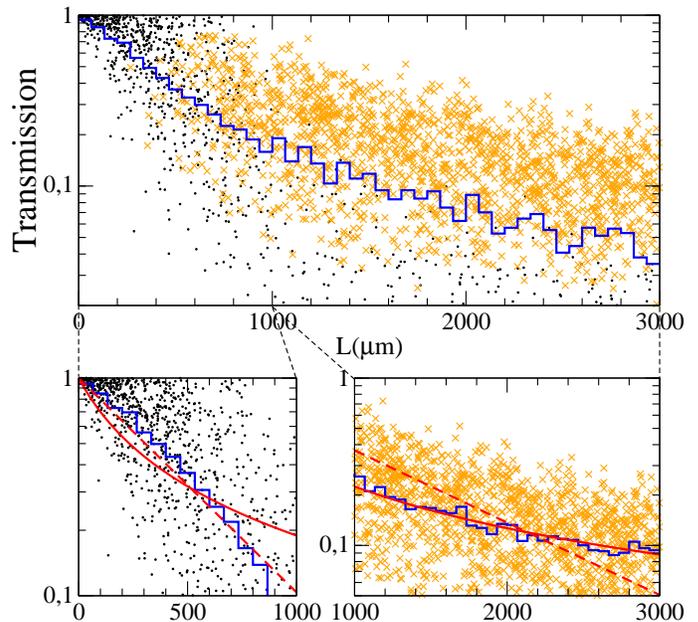}
   \caption{\label{fig10} (Color online) 
   Transition from time-independent to time-dependent
   behavior in presence of a very weak nonlinearity ($E_{int}/E_{kin}=1/100$).  
   The staircase function in the upper panel shows the geometric average of 
   both time-independent and time-dependent transmissions (black dots and orange crosses,
   repsectively).
   The lower panels display separate averages over the time-independent transmissions 
   (lower left panel) and the time-dependent transmissions (lower right panel) and
   show the best exponential (dashed line, colored red online) as well as algebraic
   (solid line, colored red online) fits to the data.
   Clearly, the time-independent transmissions decrease exponentially with $L$
   (with localization length $L_{loc}=439\, \mu$m) while
   the averaged transmissions in the time-dependent case decay according to Ohm's law.
}   
\end{figure}
The above presented computations demonstrate that even a moderate nonlinearity
leads to a dramatic change of the transmission properties. In particular,
the usual interpretation of the transmission behavior in
terms of localization is put at question in the case of interacting particles.
In order to obtain deeper insight into that matter,
we redo the above computation with a
very weak nonlinearity, such that $E_{int}/E_{kin}=1/100$.
Fig.~\ref{fig10} shows that for this case the crossover from time-independent
to time-dependent behavior is shifted to larger sample lengths
($L\gsim 600\,\mu$m).
This indicates the existence of a critical nonlinearity beyond which
the system exhibits time dependence.
Indeed, preliminary studies show that for each disorder sample length $L$
there  is a critical value $g^*$ above which no stationary
scattering state can be populated, or, equivalently, for each strength of
interaction $g$, there is a critical disorder
length $L^*$ above which the flow is time dependent.
We find that $L^*$ decreases with increasing nonlinearity.
This is the reason why stationary states can be populated in short
but not in long disorder regions and why the crossover to time-dependent dynamics
in Fig.~\ref{fig10} is displaced to larger sample lengths $L$ compared with
Fig.~\ref{fig9}.
  
From  Fig.~\ref{fig10} we also see
that the Ohmic decrease of the transmission is intimately connected
to the occurrence of time-dependent dynamics. This is clearly
indicated by the fact that the time-independent data points
can be accurately fitted by an exponential law, and not by an
algebraic one \cite{attention}. We infer from this observation that
as long as stationary states are populated the system follows the
conventional scenario of localization even in the presence of
repulsive atom-atom interactions, with a smaller localization
length than in the interaction-free case. This scenario 
seems to break down as soon as the
scattering process of the condensate becomes intrinsically
time-dependent. We tentatively attribute this phenomenon to the fact
that the definition of the mean transmission involves a {\em time
average} over the propagation process (see Eq.~(\ref{2.8})).
Therefore, information about the phase coherence, which is in
principle preserved by the time-dependent Gross-Pitaevskii equation,
becomes lost in the time-averaging procedure.

\section{Conclusions}\label{sec5a}

In this article we have presented a study of transport of Bose-Einstein
condensates in presence of disorder. We introduced a one dimensional model for
the disorder potential in the case of a condensate that propagates through a 
magnetic waveguide over a
microchip. We assumed for this model that the transverse current density in the
microfabricated wire exhibits a white-noise correlation. We could show that this
yields a disorder potential that is Lorentz-like correlated along the axis of
the waveguide.

In the regime of weak disorder, a perturbative approach allowed to estimate
the deviations from perfect transmission. We found that on length scales much
smaller than a characteristic length $L_{d}$ -- which is determined by the
correlation length of the disorder potential and the healing length of the
condensate -- the transmission decreases linearly with the length $L$ of the
disorder region. The presence of a repulsive atom-atom interaction diminishes
the transmission compared to the interaction-free case. Furthermore, in the
limit of large correlation lengths ($l_{c} \gg \lambda$) we could identify a
semiclassical regime where the backscattering is exponentially suppressed and
where the condensate density mirrors the shape of the disorder potential
$V(x)$.

The numerical approach presented in Sec.~\ref{sec2} provides an access also to 
the regime of moderate and strong disorder potentials and allows to simulate a
realistic transport process. In the case of
noninteracting atoms we find clear evidences of the appearance of
localization. In presence of interaction, time-dependence of the transmissions
becomes a dominant feature of the system. We find that stationary
scattering states can then only be populated in  waveguides with rather short
disorder regions, whereas the condensate exhibits a strongly time-dependent
dynamics if we consider large sample lengths.
Our numerical calculations show that the critical length $L^*$, at which the 
crossover between the two different regimes occurs, is shifted towards
shorter sample lengths when the strength of the interaction is increased.
It remains an open problem to determine this critical
length $L^*$ analytically from the system-specific parameters which are
the incident current, the kinetic energy, the average height of the
disorder potential, and the associated correlation length.

Our numerical study was restricted to Lorentz-correlated disorder in the
atom-chip context. We expect, however, no significant differences for other
types of disorder potentials, such as speckle fields or point scatterers.
Indeed, preliminary studies on the transport of BEC in presence of randomly
placed $\delta$-like barriers reveal qualitatively the same phenomenology: A
regime of time-dependent scattering sets in beyond a critical interaction
strength (or sample length), and the transmission decreases according to an
Ohm-like law rather than to an exponential one. An important aspect that
remains unexplored, on the other hand, is the depletion of the condensate, due
to atom-atom scattering events, and the appearance of a thermal cloud that
propagates together with the condensate. This issue should be rather relevant
in the regime of time-dependent scattering, but cannot be studied with our
present approach which is based on the mean-field description of the
condensate. Since the interaction with such a thermal cloud will lead to an
additional cause for incoherent transport, we expect no
qualitative change as far as the Ohmic power-law decay of the transmission is
concerned.

The present work opens new perspectives for the study of transport in phase
coherent systems. In particular the repulsion between the atoms
leads to a behavior different from the one expected in the non interacting
and attractive cases. 
Throughout this work we have considered realistic values of
the parameters (intensity and correlation length of the potential, distance
from the guide to the microchip, incident current of the beam) describing a
BEC of $^{87}$Rb in a waveguide and hope to motivate experimental studies
testing the results presented in this work.

\section*{Acknowledgments}

It is a pleasure to thank Isabelle Bouchoule, Ignacio Cirac, Carsten Henkel,
Markus Popp and Dirk Witthaut for fruitful and inspiring discussions. We
acknowledge the financial support by the Bayerisch-Franz\"oschisches
Hochschulzentrum (BFHZ), from the Deutsche Forschungsgemeinschaft (within the
Research School GRK 638) and from CNRS and Minist\`ere de la Recherche (Grant
ACI Nanoscience 201). Laboratoire de Physique Th\'eorique et Mod\`eles
Statistiques is Unit\'e Mixte de Recherche de l'Universit\'e Paris XI et du
CNRS, UMR 8626.

\begin{appendix}
\section{}\label{sec6}

In this appendix we derive a relation between the mean transmission
$\langle T \rangle$ and the correlation function of the disorder
potential in the weak disorder limit. Taking the mean value of
Eq.~(\ref{4.2h}) gives
\begin{equation} \label{Ap2}
\langle T \rangle=1-\frac{1}{4} \left\{ \langle\left[\delta
\rho(0)\right]^2\rangle + \frac{1}{4 \kappa^2} \langle\left[\delta
\rho'(0)\right]^2\rangle \right\} \ .
\end{equation}
Therefore the
problem of computing $\langle T \rangle$ reduces to the calculation
of the averaged values
\begin{eqnarray} \label{Ap3}
\langle\left[\delta \rho(0)\right]^2\rangle=\frac{1}{N}\sum\limits_{i=1}^{N}
\left[ \delta \rho_{i}(0) \right]^2, \nonumber\\
\langle\left[\delta \rho'(0)\right]^2\rangle=\frac{1}{N}\sum\limits_{i=1}^{N}
\left[ \delta \rho'_{i}(0) \right]^2.
\end{eqnarray}
Eq.~(\ref{4.2c}) allows us to write these averages as
\begin{eqnarray} \label{Ap4}
& & \langle\left[\delta \rho(0)\right]^2\rangle =  \frac{k^4}{\kappa^2} \times \nonumber\\
& &
\times\int\limits_{0}^{L}dx\sin(2\kappa x)\int\limits_{0}^{L}dx'\sin(2 \kappa x')
\langle v(x)v(x')\rangle \; ,
\nonumber \\
& &\langle\left[\delta\rho'(0)\right]^2\rangle=
4 k^2 \times\nonumber \\
& & \times\int\limits_{0}^{L}dx\cos(2\kappa x)\int\limits_{0}^{L}dx'\cos(2
\kappa x') \langle v(x)v(x')\rangle \; ,
\end{eqnarray}
where $\langle v(x)v(x')\rangle$ is the correlation function of the
potential. We evaluate $\langle T \rangle$ for the particularly
interesting cases of a delta-correlated white-noise potential and a
Lorentzian-correlated disorder potential. In the first case the
correlation function reads
\begin{equation} \label{Ap5}
\langle v(x)v(x')\rangle = \gamma \left(\frac{2 m}{\hbar^2 k^2}\right)^2 \pi \delta(x-x'),
\end{equation}
and the integrals in Eq.~(\ref{Ap4}) can be easily evaluated by
means of integration by parts. This yields  
\begin{eqnarray} \label{Ap6}
\langle\left[\delta \rho(0)\right]^2\rangle&=&\frac{2 m^2 \gamma \pi}
{\hbar^4 \kappa^2}~L + c \nonumber\\
\langle\left[\delta \rho'(0)\right]^2\rangle&=&4 \kappa^2
\langle\left[\delta \rho(0)\right]^2\rangle,
\end{eqnarray}
where $c$ is a dimensionless constant which is of the order 
$({4 \gamma \pi m^2})/({\hbar^4 \kappa^3})$.
For sample lengths $L \gg \kappa^{-1}$ we can neglect $c$
and we keep only terms that 
scale linearly with $L$.
By using this approximation we find with
Eq.~(\ref{Ap2}) for a white-noise random potential
\begin{equation} \label{Ap7}
\langle T \rangle=1-\frac{\pi m^2 \gamma L}{\hbar^4 \kappa^2} \ .
\end{equation}
In the case of a Lorentzian correlation,
\begin{equation} \label{Ap8}
\langle v(x)v(x')\rangle = 
\gamma~\left(\frac{2 m}{\hbar^2 k^2}\right)^2~\frac{l_{c}}{l^2_{c}+(x-x')^2},
\end{equation}
an exact analytic evaluation of the integrals~(\ref{Ap4}) is not possible.
Nevertheless, in the regime $L\gg l_{c}$ and $L\gg 1/\kappa$ the inner integral
can be approximated with high accuracy by the real or respectively the
imaginary part of
\begin{equation} \label{Ap9}
\int\limits_{0}^{L}
\frac{1}{\pi}~\frac{l_{c}~e^{2 i \kappa x'}~dx'}{l^2_{c}+(x-x')^2}
\simeq e^{2 \kappa(i x -l_{c})} \left[\Theta(x)-\Theta(x-L)\right].
\end{equation}
The same integration by parts procedure as in the white noise
potential yields a result that is only modified by the occurrence of
the exponential factor $\exp(-2 \kappa l_{c})$. Therefore, in
presence of a Lorentzian correlation the mean transmission reads
\begin{equation} \label{Ap10}
\langle T \rangle=1-\frac{\pi m^2 \gamma L}{\hbar^4 \kappa^2}e^{-2
\kappa l_{c}}.
\end{equation}

\section{}\label{sec7}

In this appendix we demonstrate that the first order solution Eq.~(\ref{4.5})
can also be derived from a standard WKB - method \cite{Landau}.
Inserting the WKB ansatz $\psi(x)=\exp(i f(x))$ (where $f(x)$ is a complex-valued
function)
in the time-independent Gross-Pitaevskii Eq.~(\ref{2.7a}) gives
\begin{equation} \label{B.1}
-\frac{\hbar^2}{2 m}\left[i f'' - (f')^2  \right]+V(x)+g e^{i(f-f^*)}=\mu.
\end{equation}
The potential $V(x)$ is supposed to be small compared to
$\mu-g n_{0}$, and its correlation length is large compared to the de Broglie
wavelength.
Eq.~(\ref{B.1}) can be rewritten in terms of a selfconsistent equation
\begin{equation} \label{B.2}
f'= \sqrt{\frac{2 m}{\hbar^2}\left(\mu-V(x)-g e^{i(f-f^*)}\right)+ i f''}.
\end{equation}
which allows to compute recursively the unknown function $f(x)$.
Treating $V(x)$ as a small perturbation motivates to use a plane
wave with wavenumber $k$ and density $n_{0}$ as zeroth-order approximation
\begin{equation} \label{B.3}
f^{(0)}=k~x - i \ln\left(\sqrt{n_{0}\,}\right).
\end{equation}
Hence, by use of Eq~(\ref{B.2}), we find
\begin{equation} \label{B.4}
{f^{(1)}}'=\frac{1}{\hbar}\sqrt{2 m \left(\mu-V(x)-g n_{0}\right)}
\equiv \frac{1}{\hbar}{\tilde P}(x),
\end{equation}
and the first-order approximation of the wavefunction reads
\begin{equation} \label{B.5}
\psi^{(1)}(x)=\sqrt{n_{0}}~\exp\left(i\int\limits_{a}^{x}  \frac{1}{\hbar}
{\tilde P}(x')dx'\right).
\end{equation}
The lower bound $a$ of the integral lies in the downstream region,
where we assume the flat condensate density $n_{0}$.
Inserting the second derivative
\begin{equation} \label{B.6}
{f^{(1)}}''= -\frac{m}{\hbar~{\tilde P}(x)}~V'(x)=
\frac{1}{\hbar}{\tilde P'}(x)
\end{equation}
into the selfconsistent
Eq.~(\ref{B.2}) yields the second-order approximation
\begin{eqnarray} \label{B.7}
{f^{(2)}}'&=& \frac{{\tilde P}(x)}{\hbar^2}\sqrt{1+i\hbar{{\tilde P'}(x)}/
{({\tilde P}(x))^2  }} \nonumber \\
&\simeq&\frac{{\tilde P}(x)}{\hbar}+i\frac{{\tilde P'}(x)}{2 {\tilde P}(x)}.
\end{eqnarray}
By integrating we find the second-order approximation of the wavefunction
\begin{equation} \label{B.8}
\psi^{(2)}(x)=\sqrt{n_{0}}\sqrt{\frac{{\tilde P}(a)}{{\tilde P}(x)}}
~\exp\left(i\int\limits_{a}^{x}  \frac{1}{\hbar}
{\tilde P}(x')dx'\right).
\end{equation}
Hence, the density found with this WKB method is
\begin{equation} \label{B.9}
n(x)= \frac{{\tilde P}(a)}{{\tilde P}(x)} =\frac{n_{0}}{\sqrt{1-v(x)}}
\end{equation}
where $v(x) \equiv V(x)/(\mu - g n_{0}) = V(x)2m/(\hbar^2 k^2)$. Indeed,
Eq.~(\ref{B.9}) coincides exactly with the result~(\ref{4.5}) found in Sec.~\ref{sec4}.

\end{appendix}


\begin{thebibliography}{10}
\bibitem{bc}This has been shown in Ref.~\cite{LeboeufPavloff} to be the
appropriate boundary condition for our problem.
\bibitem{remark} All this procedure remains perfectly valid if, instead of a
uniform distribution, $\delta j_y^\nu$ is chosen to be Gaussian distributed
around zero, with a standard deviation $\sigma/\sqrt{l}$.
\bibitem{achtung} It is however important to realize that
this analysis is only valid in the limit $k\xi\gg 1$. This is due to the fact
that, only in this limit,
the total current computed form (\ref{up3},\ref{up4}) assumes the correct value
$J_t=n_0\hbar k/m$.
\bibitem{harmosc}
Eq.~(\ref{4.2b}) is the well known equation of a driven harmonic oscillator.
$\delta \rho(x)$ is computed by use of the harmonic oscillator's Green function.
\bibitem{attention} This fact is not obvious when looking at
Fig.~\ref{fig9}, because in this figure
the extent of the time dependent region is too small
to have an influence on the fit.
\end{thebibliography}
\end{document}